\documentstyle{article}

\newcommand{\ket}[1]{{|#1\rangle}}

\author{Christof Zalka\footnote{Supported by
Schweizerischer Nationalfonds and LANL} \\ 
zalka@t6-serv.lanl.gov}

\title{An Introduction to Quantum Computers}

\begin{document}
\maketitle

\begin{abstract}
This is a short introduction to quantum computers, quantum algorithms
and quantum error correcting codes. Familiarity with the principles of
quantum theory is assumed. Emphasis is put on a concise presentation
of the principles avoiding lengthy discussions.
\end{abstract}

\section{Quantum Computers}
A quantum computer is a collection of 2-level systems (qubits). Thus
the quantum computer (QC) is described by a vector in a Hilbert space
which is the tensor product of 2-dimensional Hilbert spaces. With $l$
qubits this space has dimension $2^l$. 

The state of an $l$-qubit quantum register can be written as a
superposition of the ``computational basis states''. These are the
states where each qubit is in one of its two basis states $\ket{0}$ or
$\ket{1}$. We label these basis states by the integer which they
represent in binary. Thus:

\begin{equation} \label{QCstate}
\ket{\mbox{register}}=\sum_{n=0}^{2^l-1} c_n \ket{n}
\end{equation}

To compute, a quantum computer makes a sequence of unitary
transformations. Each unitary transformation acts on a small set
of qubits by using ``exterior fields'' which can effectively be
treated classically. Thus if, e.g., a qubit is realized by the 2-level
approximation of an atom, we can induce some U(2) transformation by
applying an electromagnetic field with the right frequency for a
certain time and with a specific phase. If no such field is applied we
assume that the state of the QC doesn't change.

A special case of unitary transformations are permutations of the
basis states; e.g., a NOT which flips the $\ket{0}$ and $\ket{1}$
states of a qubit or a so-called controlled-NOT acting as follows on
the basis states of 2 qubits:

\begin{displaymath}
U_{\mbox{\small CNOT}}~ \ket{a,b} = \ket{a,a~ \mbox{XOR}~ b} \qquad a,b=0,1
\end{displaymath}

Together with the controlled-controlled-NOT (CCNOT or Toffoli gate)

\begin{displaymath}
\mbox{CCNOT}: \quad \ket{a,b,c} \quad \to 
\quad \ket{a,b,(a~ \mbox{AND}~ b)~ \mbox{XOR}~ c} \quad ,
\end{displaymath}
we get a so-called ``universal set'' of gates for what is called
``reversible computation''. In reversible computation every gate has
as many output bits as input bits and because the gate has to be
reversible (1 to 1), only permutations of the possible input states
are allowed. It is not difficult to see that with the above 3 gates
we can compute any function of a binary input, just as we can
with conventional computation. (The CCNOT can be used as AND.)

When starting with a superposition of all ``computational'' basis
states, such as (\ref{QCstate}) the quantum computer can compute a
superposition of outputs of a function for all possible $2^l$
inputs. Now, at the end of the quantum computation we have to make a
measurement on the quantum computer. We measure for each qubit whether
it is in state $\ket{0}$ or $\ket{1}$. Thereby we collapse the state
of the QC onto a computational basis state $\ket{n}$ with probability
$|c_n|^2$.

If we do this to a superposition of function values, this is not going
to be of interest. The trick is to look for interference between the
computational basis states. For this we have to add ``non-classical''
gates to the above ones, that is, gates which transform computational
basis states into a superposition thereof.

Shor's quantum algorithm for factoring large integers \cite{shor1}
first computes a superposition of functional values and then applies a
number of ``non-classical'' gates before measuring the QC. The
non-classical gates bring the QC into a state where only about square
root of the computational basis states have a sizable amplitude
(coefficient $c_n$), thus the final measurement will pick one of
them. The observed basis state $\ket{n}$ will thus have a ``random
component'' but also carries some information which can be used to
solve the mathematical problem at hand. Shor's algorithm is described
in more detail in section \ref{shor}.

\section{Possible technical realizations: ions in an electromagnetic trap}

One proposal \cite{cirac} for building a quantum computer is to use a
linear ion trap. In such a trap a number of ions will line up along
the $z$-axis. Along the $x$- and $y$-axis they are strongly confined
by a high-frequency electric field that switches between being
confining in the $x$-direction and deconfining in the $y$-direction
and vice versa. The net effect is that the ions are confined strongly
in both directions. In the $z$-direction the ions are confined by a
relatively weak harmonic electric potential. Due to their mutual
electrostatic repulsion the ions will form a string with separations
of the order of micro meters.

Each ion represents a qubit. By shining laser light at an individual
ion, U(2) transformations can be applied to that qubit. For
``universal quantum computation'' we at least also need to be able to
induce unitary transformations on a pair of ions such that the initial
product states will become non-product states (``entangled'' states).

To do this, the ions have to be cooled so that they occupy the lowest
energy state of their motional degrees of freedom in the confining
potential.
It is not so difficult to do this for the strongly confining $x$- and
$y$-directions. For the $z$-direction, instead of looking at the
individual motions of the ions, one looks at collective motions,
``normal modes'' which are like uncoupled harmonic oscillators. The
lowest such mode is the ``center-of-mass'' mode where the ions can be
imagined to oscillate synchronously without changing their spacings.

The idea now is to use the two lowest states of this center-of-mass
oscillator as a ``bus-qubit''. The internal state of an ion can be
coupled to the ``center-of-mass'' degree of freedom, e.g., by shining
laser light at it with a frequency which will take the ground state of
the ion and the center-of-mass motion to the first excited state of
both.

This has been done experimentally with a single ion. Also pictures of
some 30 ions aligned in a linear ion trap (and fluorescing in laser
light) have been obtained, but presently the main problem is to cool
such a string of ions to the motional ground state. One tries to
achieve this with laser-cooling (doppler-cooling and others). See e.g.
\cite{monroe}.

\section{Quantum algorithms}

\subsection{factoring large integers} \label{shor}

The quantum factoring algorithm arguably is the only known case where
a quantum computer could solve an interesting problem much faster than
a conventional computer. Actually the computation time is a small
power ($2^{nd}$ or $3^{rd}$) of the number of digits of the integer to
be factored. The fastest known classical integer-factoring algorithms
use super-polynomial time, and it is believed that no polynomial-time
such algorithms exist.

Shor's quantum factoring algorithm relies on the fact that factoring
can be ``reduced'' to the problem of finding the period of the
following periodic function:

\begin{displaymath}
 f(x)=a^x~ \mbox{mod}~ N \quad ,
\end{displaymath}
where $N=p\cdot q$ is the number to be factored and $a$ is essentially
an arbitrary constant. Note that Euler's generalization of Fermat's
little theorem states that \mbox{$a^{(p-1)(q-1)} \mbox{mod} (pq)
=1$}. Thus $(p-1)(q-1)$ is a multiple of the period of $f(x)$. It
should therefore be plausible that there are efficient ways to get the
factors $p$ and $q$ from the period.

After computing a superposition of the form $\sum \ket{x,f(x)}$, the
period can be found by employing the ``quantum Fourier transform''
which applies the discrete Fourier transform to the $2^l$
amplitudes of a quantum register.

To obtain $\ket{x,f(x)}$ with reversible computation we have to make a
little detour. By using additional qubits in the state 0, we can use
CNOT and CCNOT to compute XOR and AND, but we also produce unwanted
output bits. In a superposition, the unwanted qubits will be
quantum-correlated (entangled) with the wanted qubits. Observing and
resetting the unwanted qubits doesn't work, as we thereby also
collapse the wanted part of the QC. The trick is to first compute
$f(x)$ including the garbage $g(x)$, then copy $f(x)$ into a ``save''
register and then undo the first step, which is of course possible in
reversible computation:

\begin{displaymath}
\ket{x,0,0,0} ~\to~ \ket{x,f(x),g(x),0} ~\to~ \ket{x,f(x),g(x),f(x)} 
~\to~ \ket{x,0,0,f(x)}
\end{displaymath}

The copying of $f(x)$ in the second step can be done with a sequence of
CNOT's.

So by starting with a superposition with equal amplitudes of all $x$ ,
we get:

\begin{displaymath}
\frac{1}{\sqrt{2^l}} \sum_{n=0}^{2^l-1} \ket{x,0} \quad \to \quad 
\frac{1}{\sqrt{2^l}} \sum \ket{x,a^x \bmod N}
\end{displaymath}

For the following it is easier to imagine that now we measure the
second register, but it is not necessary. After such a measurement the
first register will be in a superposition of all $x$ that give the
measured output value. The value of the smallest such $x$ is random,
but the spacing of the following values is the period of $f(x)$ which
we want to know. Thus the amplitudes in the first register are peaked
with constant spacings between the peaks, but everything is shifted by
a random value. When applying the quantum Fourier transform to the
first register we will again get amplitudes that are peaked at regular
intervals, but now the first peak is at the origin and the random
shift in the previous peaks only shows up as some complex phase of the
peaks. (To get sharp peaks we choose the size of the $x$-register such
that $2^l$ is at least of the order of the period squared.) By
measuring the Fourier transformed register, and repeating the whole
quantum computation a few times, one obtains the spacing of the peaks.

The quantum Fourier transform is done by using the fast Fourier
transform algorithm (FFT), which applies very naturally to
transforming the amplitudes of a quantum register. Actually it can be
done so efficiently that it is negligible for the overall computation
time.

\subsection{unstructured search}

To search through $N$ cases, there is a simple quantum algorithm
\cite{grover} that takes some $\sqrt{N}$ steps. This is not a very
strong improvement over the classical case with $N$ steps. Furthermore
it can be shown that for this problem no better quantum algorithm
exists \cite{bennett, zalka3}. To prove this, the unstructured search
is formalized with a so-called ``oracle'', a black-box subroutine
which gives output 1 only for one out of all possible inputs.

\subsection{simulating arbitrary quantum systems}

The amplitudes of the computational basis states of a QC can be made
to follow the time evolution of the amplitudes of essentially any
quantum system \cite{zalka1, wiesner}. Obtaining information about the
quantum state is then of course restricted by the same fundamental
quantum principles as it is for the original quantum system.

Say we have a quantum mechanical Hamilton operator that is a sum of
a kinetic and a potential term, thus a sum of a term which is a function
of momentum operators and a term that is a function of position
operators. We now discretize the wave function of this quantum
mechanical system and ``store'' it as the amplitudes of the
computational basis states of the QC. The point is that we can go to
momentum space by applying the quantum Fourier transform. Then we will
have the discretized wavefunction in momentum space. Time evolution is
implemented by evolving the wavefunction for a short time only
according to the potential term, then go to momentum space and evolve
according to the kinetic term, and so on.

Evolving the wavefunction for a short time according to e.g. only the
potential term, amounts to multiplying with a complex phase. We have
to carry out a transformation of the form $\ket{x} \to e^{i f(x)}
\ket{x}$, which can be done quite easily.

\section{Quantum error correcting codes}

To carry out a long quantum computation seems to require very precise
operations and low noise. From imprecisions in the applications of the
exterior fields (lasers, etc.) quantum gates will always be somewhat
different from the intended unitary operations. In this respect we have the
same problems as with an analog computer where, contrary to digital
hardware, slightly deviating values are not automatically reset to a
standard value. Also it is very difficult to isolate the degrees of
freedom of a quantum computer from the environment.

Therefore quantum computation might well be practically impossible,
were there not the possibility of quantum error correction.
Interestingly, it is possible to correct (continuous) quantum
amplitudes much better than e.g. continuous classical quantities in an
analog computer.

To simplify things, the following discussion will not encompass the
most general possibilities for quantum error correction. I will
describe how the 5-qubit error correcting code \cite{cesar} works, which
has been shown to be the shortest possible quantum code.

So we want to encode (and thus protect) a qubit in 5 qubits. The code
is a 2 dimensional subspace of the 32 dimensional Hilbert space of the
code. How can we correct for errors? Making a full (no degenerate
eigenvalues) measurement of the code in order to correct for errors is
not good as this will collapse the (generally unknown) encoded
qubit. The trick is to only measure what error has affected the code
without learning anything about the encoded qubit. The eigenspaces of
the error measurement have to be at least 2 dimensional so that the
encoded quantum information will not collapse.

An error correcting code can of course only correct for some errors,
ideally the most probable ones. The standard assumption is that the
probabilities of errors affecting different physical qubits are
independent. Then it makes sense to correct for all errors which
affect just one qubit.

For every qubit there are 3 such errors, namely bit flips, phase flips
and bit-phase flips, corresponding to the 3 Pauli matrices. We must
also take into account that no error may have happened. Then for the
5-qubit code we have $5 \cdot 3 +1$ possible ``errors''. After having
been exposed to noise the code will in general be in a superposition
of the 16 resulting states, but a measurement of the error will
collapse it to one of these states. Thus for convenience we can simply
imagine that one of the 16 errors has happened.

Now let's see what conditions the code (= the 2-dim. code subspace)
has to fulfill. The condition is that the 16 images under the error
operators of the 2-dim. code subspace have to be pairwise
orthogonal. This makes it possible to construct an ``error
observable'' with these eigenspaces. Note that the 16 mutually
orthogonal 2-dim. subspaces ``fill'' the 32 dimensional code Hilbert
space, which is why the 5 qubit code is called ``perfect''.

Once we have determined which error has happened we simply undo this
error (it's a unitary transformation). It may technically not be
feasible to directly measure the error observable, but one can apply a
series of quantum gates such that the error observable then
corresponds to simply measuring 4 of the 5 qubits.

Such codes can, e.g., be used to store an (unknown) quantum state or
transmit it through a noisy transmission line. In quantum computing we
also must be concerned about the noisiness of the error correction
operations. Also, we never want to decode the processed quantum
information, as this would expose it to noise. Schemes have been
developed for ``fault tolerant quantum computing'' where error
correction operations can increase the probability of having an
undisturbed state even though they are noisy themselves, and
operations (quantum gates) can be applied to encoded qubits such that
again some level of noise can be tolerated. If we imagine one of our
standard errors $E_i$ to happen suddenly with some probability, then
such schemes can tolerate a certain number of such errors before the
quantum computation goes wrong, provided that the errors don't
cluster too much. Without fault tolerant quantum computing, a single
error would already be too much.

One can think of iterating the encoding procedure, thus, e.g., one
qubit would be encoded in five qubits which in turn would each be
encoded in five qubits. It has been shown that with such a scheme
arbitrarily long quantum computations could be carried out once the
rate of errors per physical qubit is below a certain level of the
order of $10^{-4}$ per operation (see e.g. \cite{zalka2}).

\section{Recommended papers}

{\bf introductions:} \\
\cite{cipra} is a semi-popular short introduction. \\ 
Also John Preskill gives a lecture on quantum computation at Caltech. 
The Web site \cite{preskill} contains the lecture notes and other 
useful information and links. \\ \\
{\bf factoring:} \\
Shor's paper \cite{shor1} is well written but long. A short account of 
Shor's algorithm is also given in \cite{laf}. \\ \\
{\bf error correction:} \\ 
In \cite{shor3} Shor describes how his 9-qubit quantum error
correcting code works. This was the first quantum code. \\ \\
{\bf fault tolerant quantum computing:} \\
In \cite{shor2} Shor describes how to carry out fault tolerant error
correction and fault tolerant operations on the 7-qubit and related
codes.

\end{document}